\documentclass[11pt]{article}
\usepackage{moriond,epsfig}

\def\Journal#1#2#3#4{{#1} {\bf #2}, #3 (#4)}

\def\PLB{{\em Phys. Lett.}  B}
\def\PRL{\em Phys. Rev. Lett.}
\def\PRD{{\em Phys. Rev.} D}

\def\EPJ{{\em Eur. Phys. J.} C}

\def\be{\begin{equation}}
\def\ee{\end{equation}}
\def\bea{\begin{eqnarray}}
\def\eea{\end{eqnarray}}

\begin{document}
\vspace*{4cm}
\title{OPEN AND HIDDEN CHARM PRODUCTION WITH THE HERA-B EXPERIMENT}

\author{ A. BOGATYREV }

\address{Institute of Theoretical and Experimental Physics, 117259 Moscow, Russia}

\maketitle\abstracts{
Measurements of the suppression of the yield per nucleon and differential distributions of $J / \psi$ production for 920 GeV/c protons incident on heavy nuclear targets have been made with broad coverage in $p_T$ and negative coverage in $x_F$ of produced meson. Production ratios of $\psi(2S)$ to $J / \psi$ and $\chi_c$ to $J / \psi$ have been measured with a high accuracy. The $D^+$ and $D^0$ production cross sections as well as $D^+$ to $D^0$ ratio have been obtained on one of the highest statistics available in proton nucleus experiments.}

\section{Introduction}
The mechanisms by which charmonium and open charm states are produced in hadronic collisions are not well understood and are the focus of much interest. 
Although most of the results from previous experiments are qualitatively described by different theoretical models, there are some quantitative discrepancies that may be resolved with a better understanding of non-pertubative effects in the production of heavy quarks.
Taking into account that the production mechanisms can be identified by their strong kinematic dependences, it is crucial to have new measurements with high statistics and broad kinematic coverage to accomplish comprehensive descriptions of charm production in nuclei. In this context the HERA-B experiment provides new data on open and hidden charm production. 
\section{The HERA-B Experiment}
HERA-B is a fixed target experiment at the HERA storage ring at DESY. Diverse hidden and open charm states are produced in proton-nucleus inelastic collisions. The $pN$ energy in the center of mass corresponds to 41.6 GeV. 

The spectrometer emphasizes vertexing, tracking, particle identification and features a dedicated multilevel $J / \psi$ trigger.  

The HERA-B target is a fixed target system consisting of two stations separated by 4 cm along the beam direction. Each station houses 4 target wires which can be moved independently into the beam halo. 

\begin{figure} 
\begin{center}
\psfig{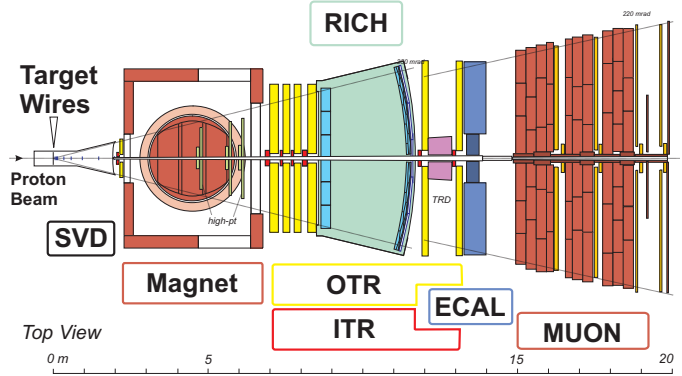}
\end{center}
\caption{Layout of the HERA-B detector.
\label{fig:det}}
\end{figure}

The side view of the detector is shown in Fig.~\ref{fig:det}. The protons come from the left and interact in the target region. The target station is followed by the Vertex Detector System(VDS), which provides a precise reconstruction of primary and secondary vertices. The main tracking system(ITR+OTR) is placed behind the VDS. It uses different technologies, depending on the distance to the beam center.

Particle identification is performed by a Ring Imaging Cherenkov Counter(RICH), an Electromagnetic Calorimeter(ECAL) and a MUON detector(MUON).

The $J / \psi$ trigger~\cite{herabtrigger} is initiated by di-lepton signatures from either MUON or ECAL systems. The information on regions of interest defined by pretriggers is used as seeds for the following search for tracks in the tracking system. Then the trigger performs the extrapolation of found tracks through the magnet and follows them through the VDS. In addition a vertex constraint is applied. 
Events with at least two fully reconstructed di-lepton candidates which come from a common vertex pass the trigger requirements.    
\section{Data Set 2002/3}
The HERA-B collaboration collected data during 3 month data-taking period in 2002/2003. The data samples were obtained using carbon, titanium and tungsten wires. During that time 150 million di-lepton triggered events were acquired. About 170,000 of $J / \psi$ were reconstructed in muon channel and about 150,000 of $J / \psi$ were found in electron channel. The trigger performance has been significantly improved since the year 2000. For comparison, in 2000 the $J / \psi$ rate was 25-35 per hour and in 2002/2003 we already had 1000-1500 $J / \psi$ per hour. 
Significant improvements in the data acquisition system had been made as well, which resulted in achieving high recording rates, which in turn allowed accumulation of 210 million minimum bias events in just one week of running. The Minimum Bias trigger selected events satisfying a requirement on either a minimum number of hits in RICH or a minimum energy deposited in ECAL. It allowed to suppress events with no interaction without introducing any significant bias. 

\section{Preliminary results. The Minimum Bias Data}

The following open charm decays are fully reconstructed in minimum bias data:

$D^0 \rightarrow K^- \pi^+$

$D^+ \rightarrow K^- \pi^+ \pi^+$

$D^{*+} \rightarrow D^0 \pi^+$ \hskip 0.4cm $D^0 \rightarrow K^- \pi^+$ \\
Throughout this paper charge conjugate decays are also implied.

\begin{figure}[h]
\begin{center}
\begin{minipage}{2.0in}
\begin{center}
\psfig{figure=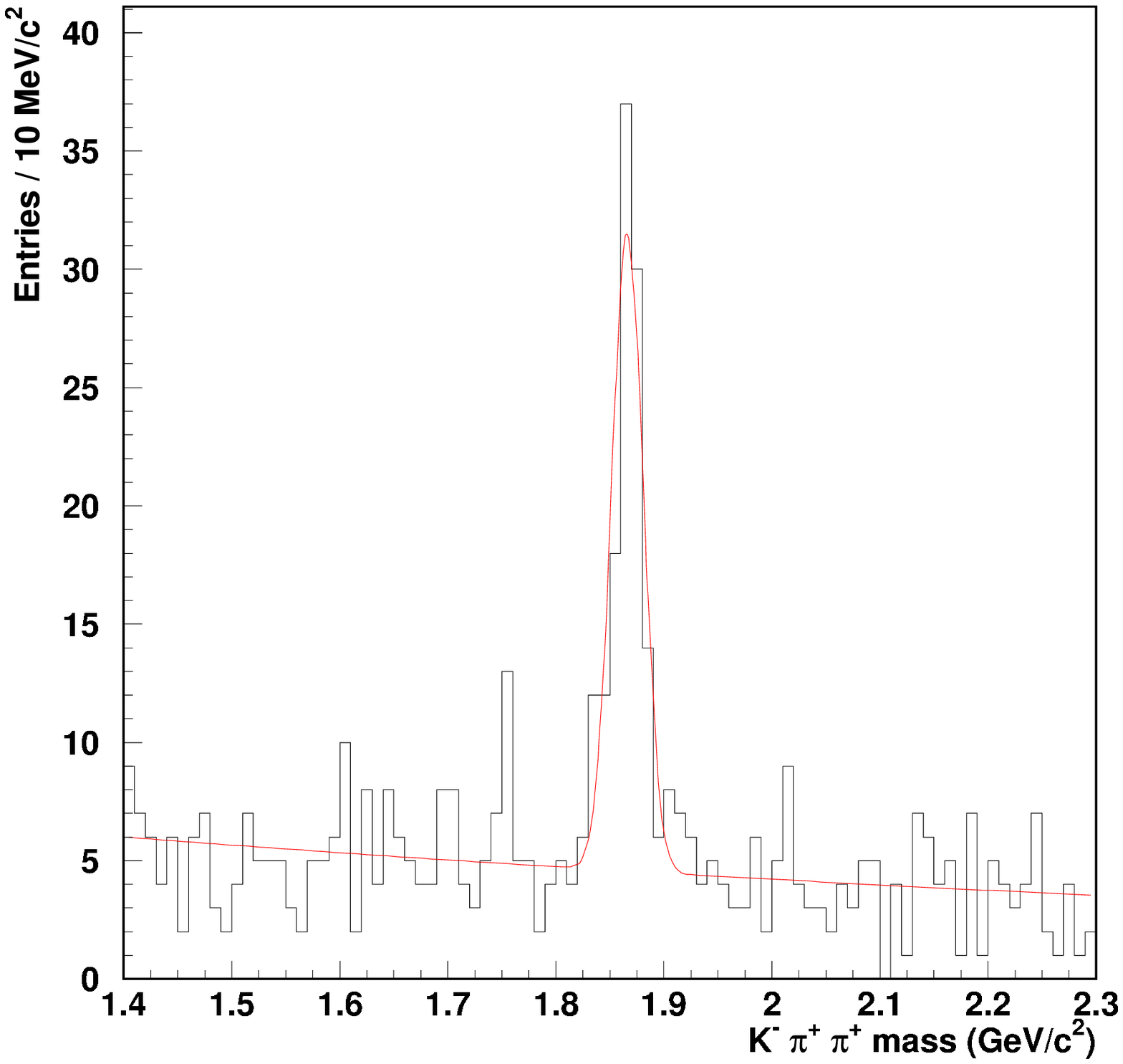,height=1.8in}
\end{center}
\caption{$M(K\pi\pi)$.
\label{fig:dplus}}
\end{minipage}
\begin{minipage}{2.0in}
\begin{center}
\psfig{figure=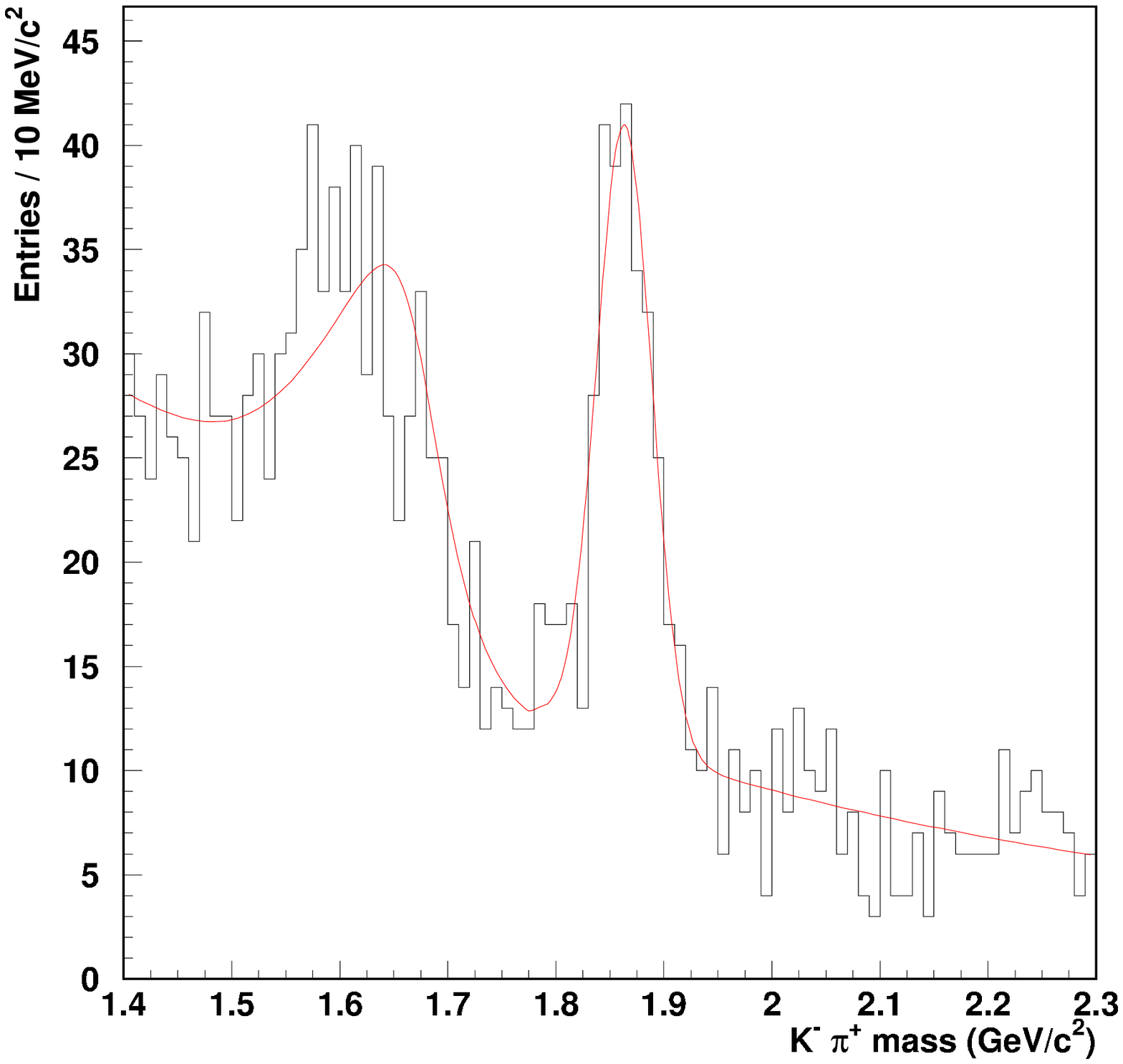,height=1.8in}
\end{center}
\caption{$M(K\pi)$.
\label{fig:d0}}
\end{minipage}
\begin{minipage}{2.0in}
\begin{center}
\psfig{figure=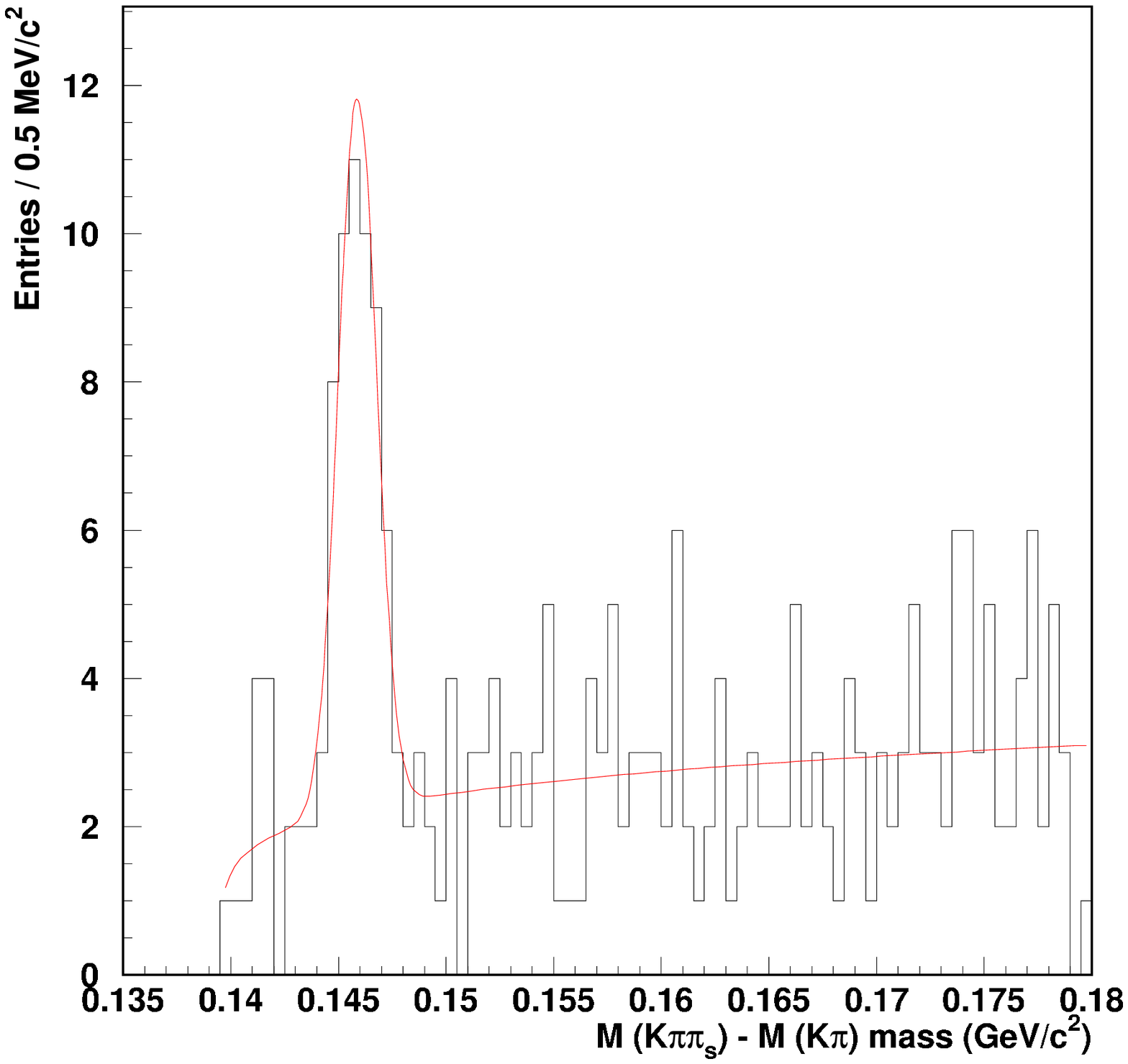,height=1.8in}
\end{center}
\caption{$M(K\pi\pi_{slow})-M(K\pi)$.
\label{fig:dstar}}
\end{minipage}
\end{center}
\end{figure}

The invariant mass plots for $D^+$ and $D^0$ are shown in Fig.~\ref{fig:dplus} and Fig.~\ref{fig:d0} respectively. 
The bump seen to the left of the $D^0$ in Fig.~\ref{fig:d0} is produced by reflection of not fully reconstructed charm decays, for instance $D^0 \rightarrow K^- \pi^+ \pi^0$.  
The shape of this reflection was obtained from MC generated $c\overline{c}$ events and was used in the fit as a component of the background in $K \pi$ invariant mass plot. 

The mass difference between the mass of $D^{*+}$ and the mass of $D^0$ is shown in Fig.~\ref{fig:dstar}. This quantity peaks close to the pion mass and has a very small width. The peak is narrow due to the fact that the errors in calculation of the invariant mass of three particles (in case of $D^{*+}$) and two particles (in case of $D^0$) are highly correlated, and therefore cancel out to a large extent. 
 
The signal yields for the fully reconstructed $D^+$, $D^0$ and $D^{*+}$ decays are shown in Table~\ref{tab:dyields}.

\begin{table}[ht]
\caption{Signal yields for $D^0$, $D^+$ and $D^{*+}$.\label{tab:dyields}}
\vspace{0.4cm}
\begin{center}
\begin{tabular}{|c|c|c|}
\hline
$D^+ \rightarrow K^- \pi^+ \pi^+$ &
$D^0 \rightarrow K^- \pi^+$ &
$D^{*+} \rightarrow D^0 \pi^+$  
\\ \hline
$98\pm12$ &
$189\pm20$ &
$43\pm8$
\\ \hline
\end{tabular}
\end{center}
\end{table}

The sample of open charm collected by HERA-B constitutes one of the highest statistics available in proton beam experiments. Therefore the HERA-B will provide the precise measurement of $D^0$, $D^+$ and $D^{*+}$ cross sections as well as $\frac{D^+}{D^0}$ and $\frac{D^{*+}}{D^0}$ ratios. 

\begin{table}[ht]
\caption{Results on $D^+$ and $D^0$ production cross sections.\label{tab:dcross}}
\vspace{0.4cm}
\begin{center}
\begin{tabular}{|c|c|c|}
\hline
&
$-0.1<x_F<0.05$ &
full $x_F$ 
\\ \hline
$ \sigma(D^+)\mu b/nucl$ &
$11.5 \pm 1.7_{stat} \pm 2.2_{sys}$ &
$30.2 \pm 4.5_{stat} \pm 5.8_{sys}$
\\ \hline
$ \sigma(D^0)\mu b/nucl$ &
$21.4 \pm 3.2_{stat} \pm 3.6_{sys}$ &
$56.3 \pm 8.5_{stat} \pm 9.5_{sys}$
\\ \hline
\end{tabular}
\end{center}
\end{table}

The preliminary results on $D^+$ and $D^0$ cross sections are summarized in Table~\ref{tab:dcross}. The $x_F$ range of this analysis extends from -0.1 to 0.05. Also shown are the total cross sections which were calculated by extrapolating over the full $x_F$ range. 
In the calculation of the ratio between $D^+$ and $D^0$ cross sections the systematic errors related to detector efficiencies are minimized and the dependence on the absolute luminosity determination is removed. This allows the accurate measurement of the cross section ratio. Result for the $\frac{D^+}{D^0}$ ratio is 
$0.54 \pm 0.11_{stat} \pm 0.14_{sys}$. 

\begin{figure}[ht]
\begin{center}
\begin{minipage}{2.1in}
\begin{center}
\psfig{figure=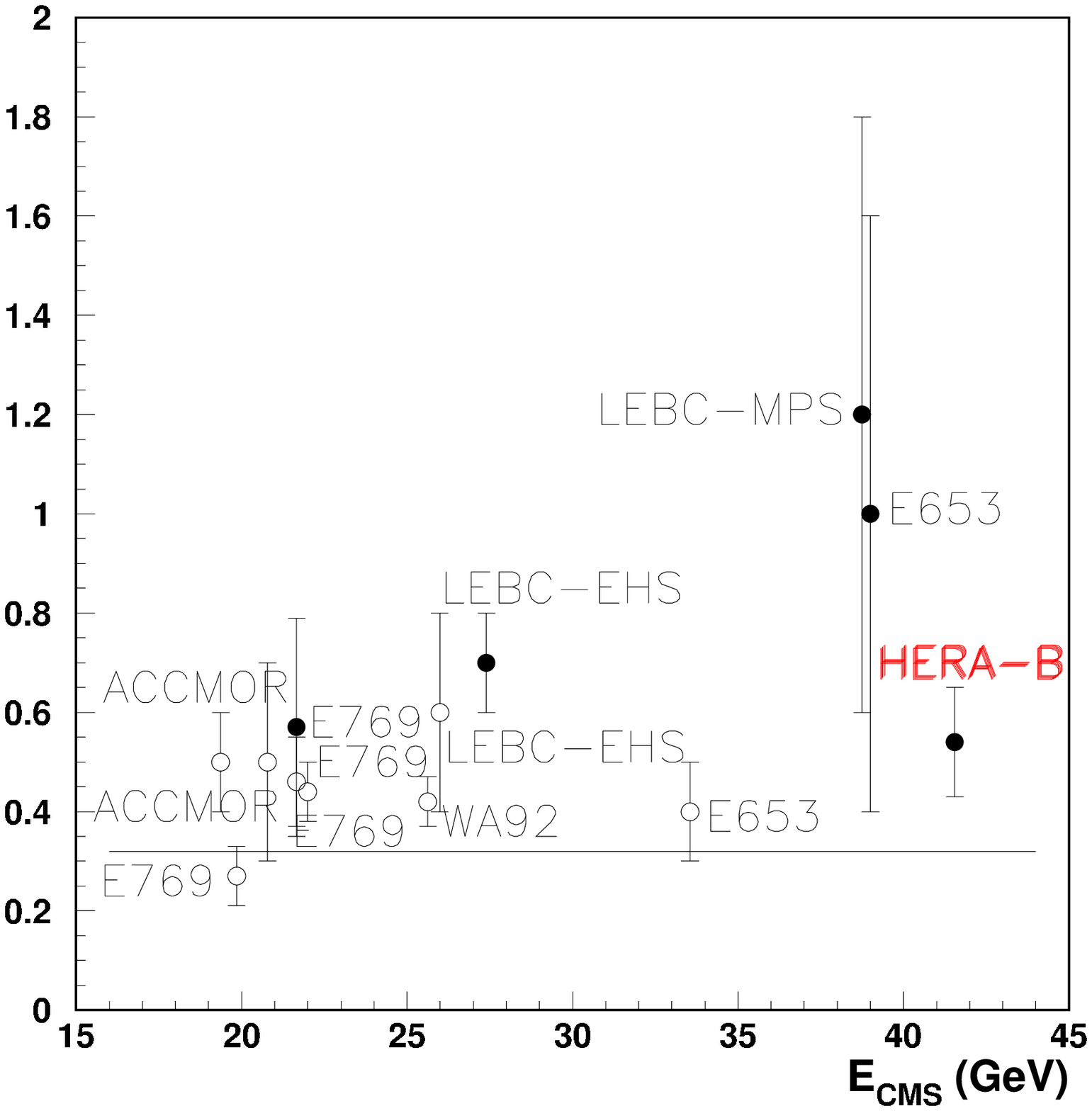,height=2.1in}
\put(-130,135){$\frac{D^+}{D^0}$}
\end{center}
\caption{Ratio between $D^+$ and $D^0$ production cross sections. The line shows theoretical expectation.
\label{fig:wratio}}
\end{minipage}
\hspace*{20mm}
\begin{minipage}{2.1in}
\begin{center}
\psfig{figure=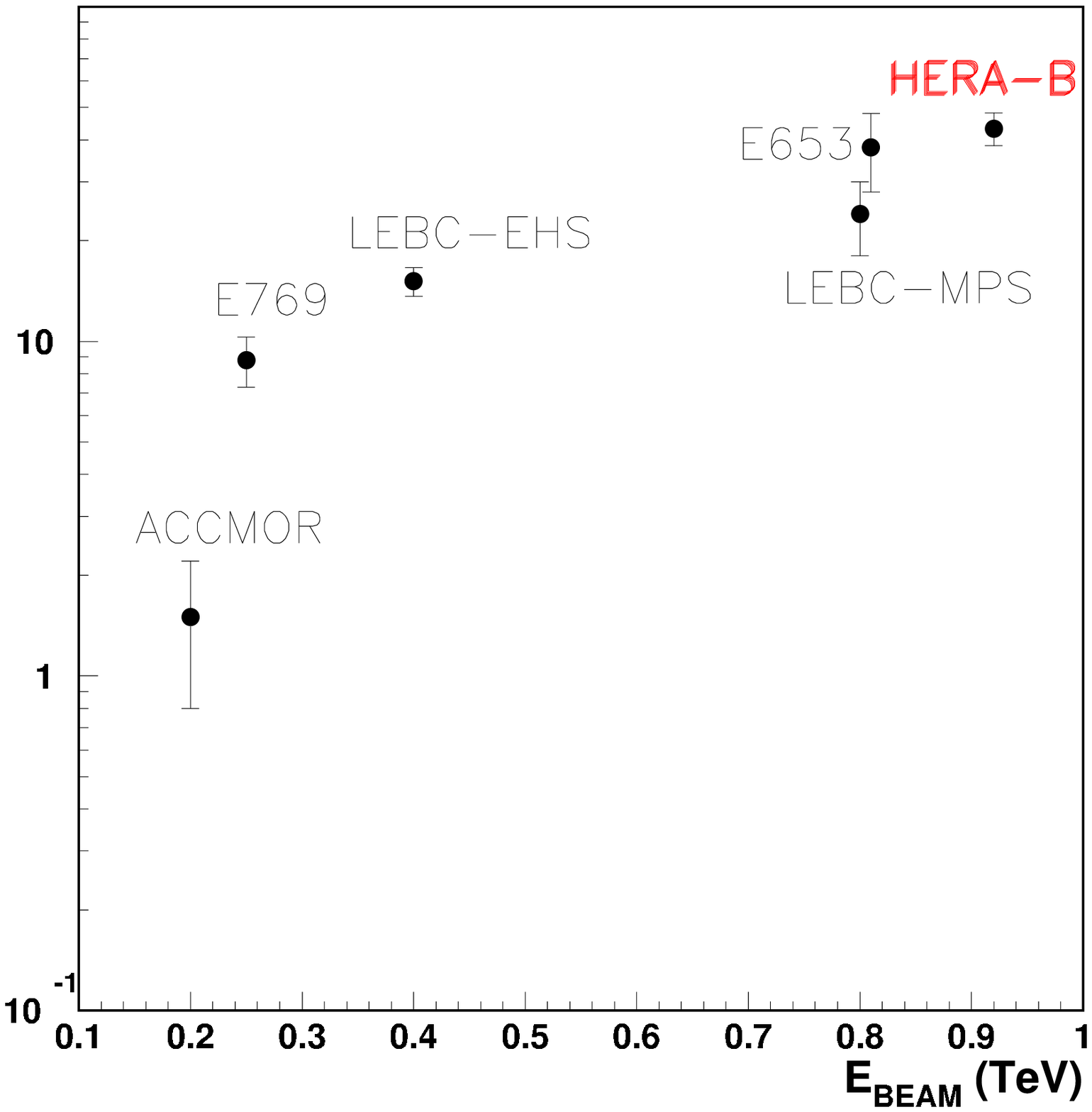,height=2.1in}
\put(-133,137){$D\overline{D}$($\mu b/nucl$)}
\end{center}
\caption{Comparison of our $D\overline{D}$ cross section with the results from previous experiments.
\label{fig:cross}}
\end{minipage}
\end{center}
\end{figure}

The comparison of our measurement for $\frac{D^+}{D^0}$ ratio with those of other experiments is shown in Fig.~\ref{fig:wratio}.
For the ratio the experimental situation is unclear, and the uncertainties are large particularly for proton beam experiments. The line on the plot shows the expected theoretical value for the ratio of charged to neutral $D$ mesons which is approximately equal to $\frac{1}{3}$ and is the same for the proton and pion induced interactions~\cite{frix}. The big difference between the $D^+$ and $D^0$ production results from the fact that $D$ mesons can be produced either directly or through feed-down 
from $D^*$ decays. The prediction for the ratio is obtained by taking into account the well known branching ratios for the decays of $D^*$ into $D$ mesons.

Our result on $D\overline{D}$ cross section is shown in Fig.~\ref{fig:cross}. It is compatible with previous data and has smaller experimental uncertainty in comparison with the experiments at nearby proton beam energies.

\section{Preliminary results. The Di-lepton Data}
\subsection{The Di-lepton invariant mass spectra}\label{subsec:prod}
\begin{figure}[ht]
\begin{center}
\hspace*{-18mm}
\begin{minipage}{2.3in}
\begin{center}
\psfig{figure=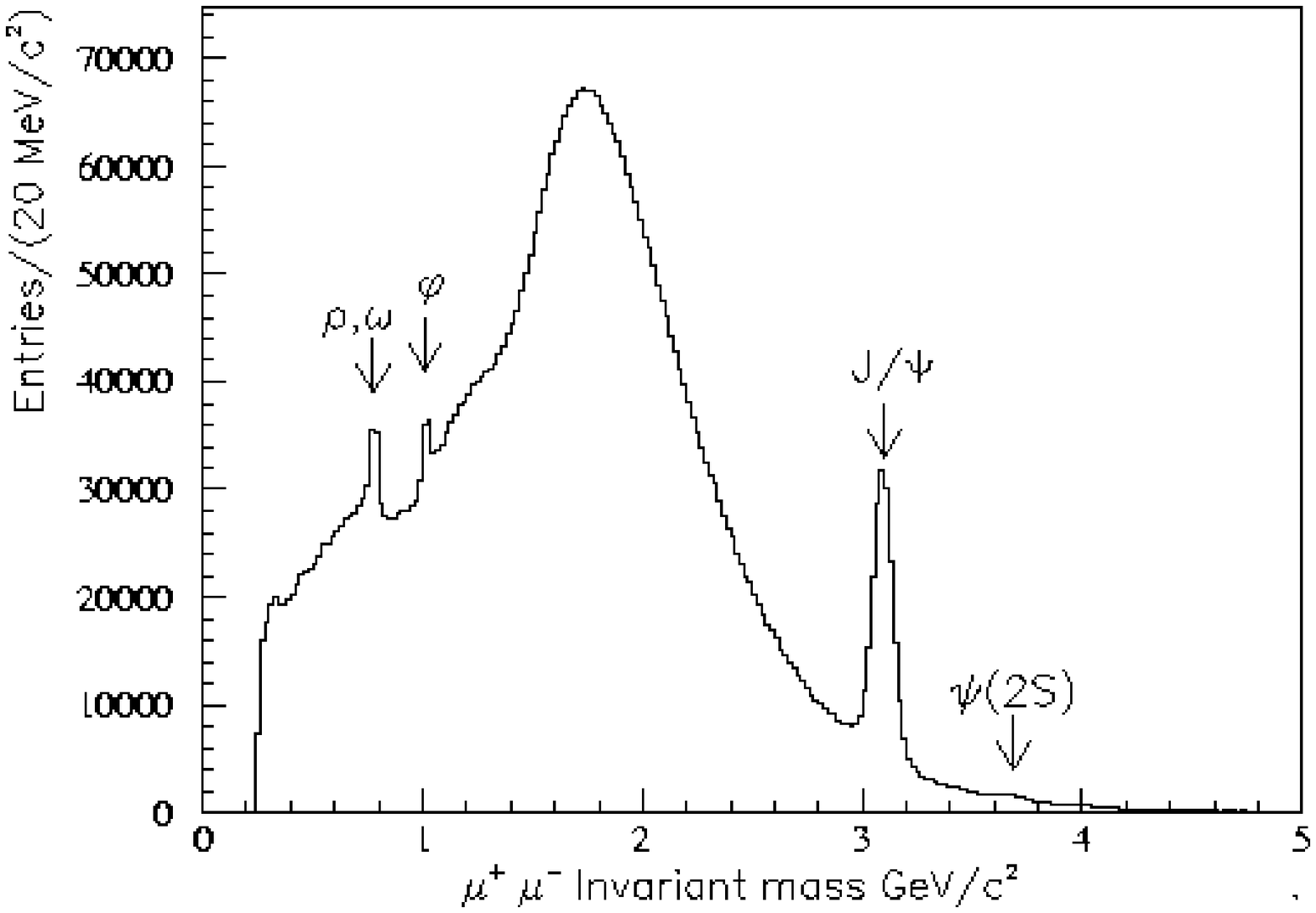,height=2.05in,,width=2.91in}
\end{center}
\caption{$\mu^+\mu^-$ invariant mass spectrum. Di-lepton trigger.
\label{fig:jpsimumu}}
\end{minipage}
\hspace*{18mm}
\begin{minipage}{2.3in}
\begin{center}
\psfig{figure=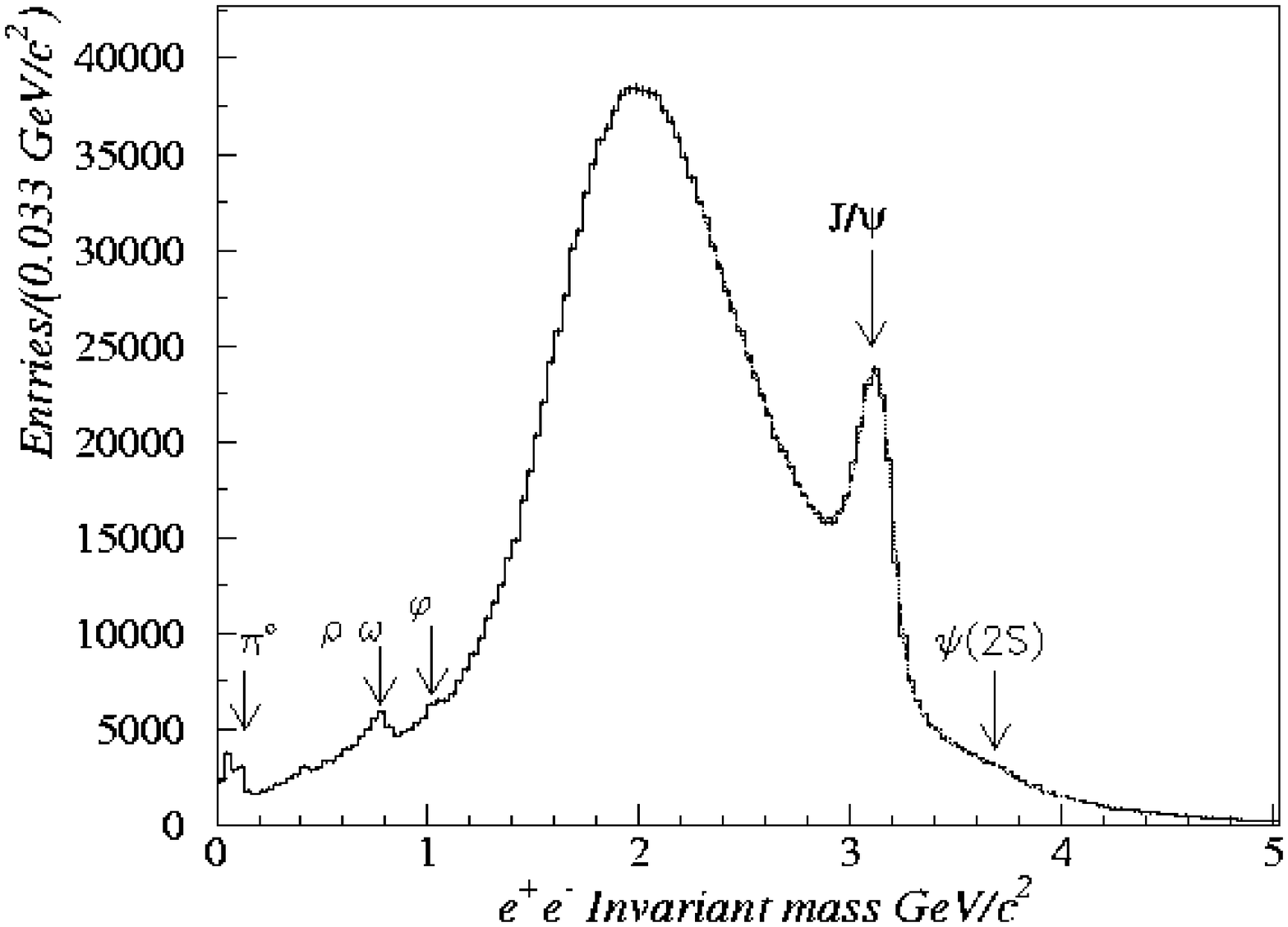,height=2.1in}
\end{center}
\caption{$e^+e^-$ invariant mass spectrum. Di-lepton trigger.
\label{fig:jpsiee}}
\end{minipage}
\end{center}
\end{figure}

The $\mu^+\mu^-$ and $e^+e^-$ invariant mass spectra are shown in Fig.~\ref{fig:jpsimumu} and Fig.~\ref{fig:jpsiee} respectively. Clear $J / \psi$ and $\psi(2S)$ resonance peaks are visible. Lower mass particle decays of $\rho$, $\omega$, $\phi$ into $l^+l^-$ pair are also seen.

\subsection{$J / \psi$ Differential Distributions}\label{subsec:prod}
\begin{figure}[ht]
\begin{center}
\hspace*{-2mm}
\begin{minipage}{2.75in}
\begin{center}
\psfig{figure=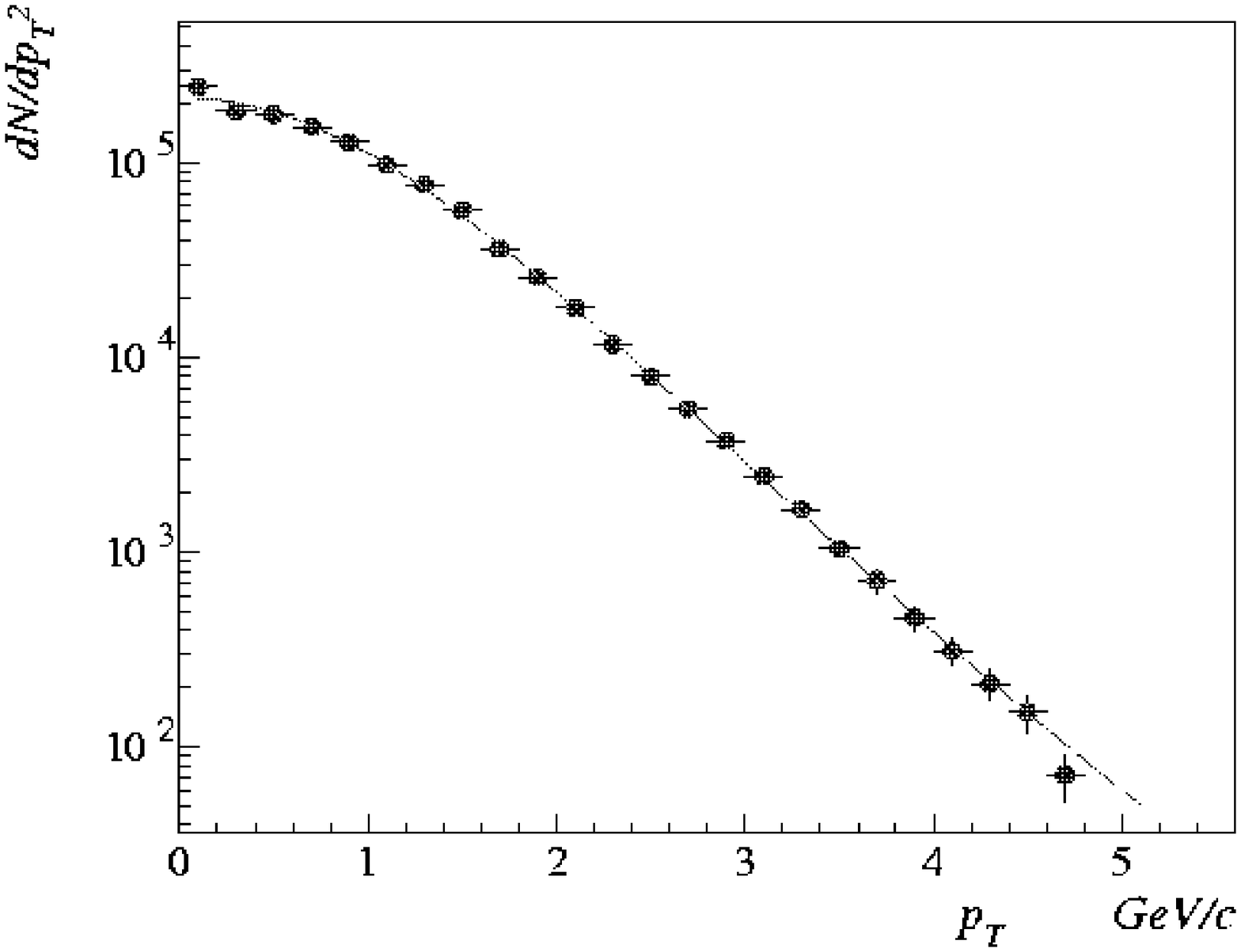,height=2.0in}
\end{center}
\caption{$J / \psi$ differential distribution versus $p_T$.
\label{fig:pt}}
\end{minipage}
\hspace*{5mm}
\begin{minipage}{2.75in}
\begin{center}
\psfig{figure=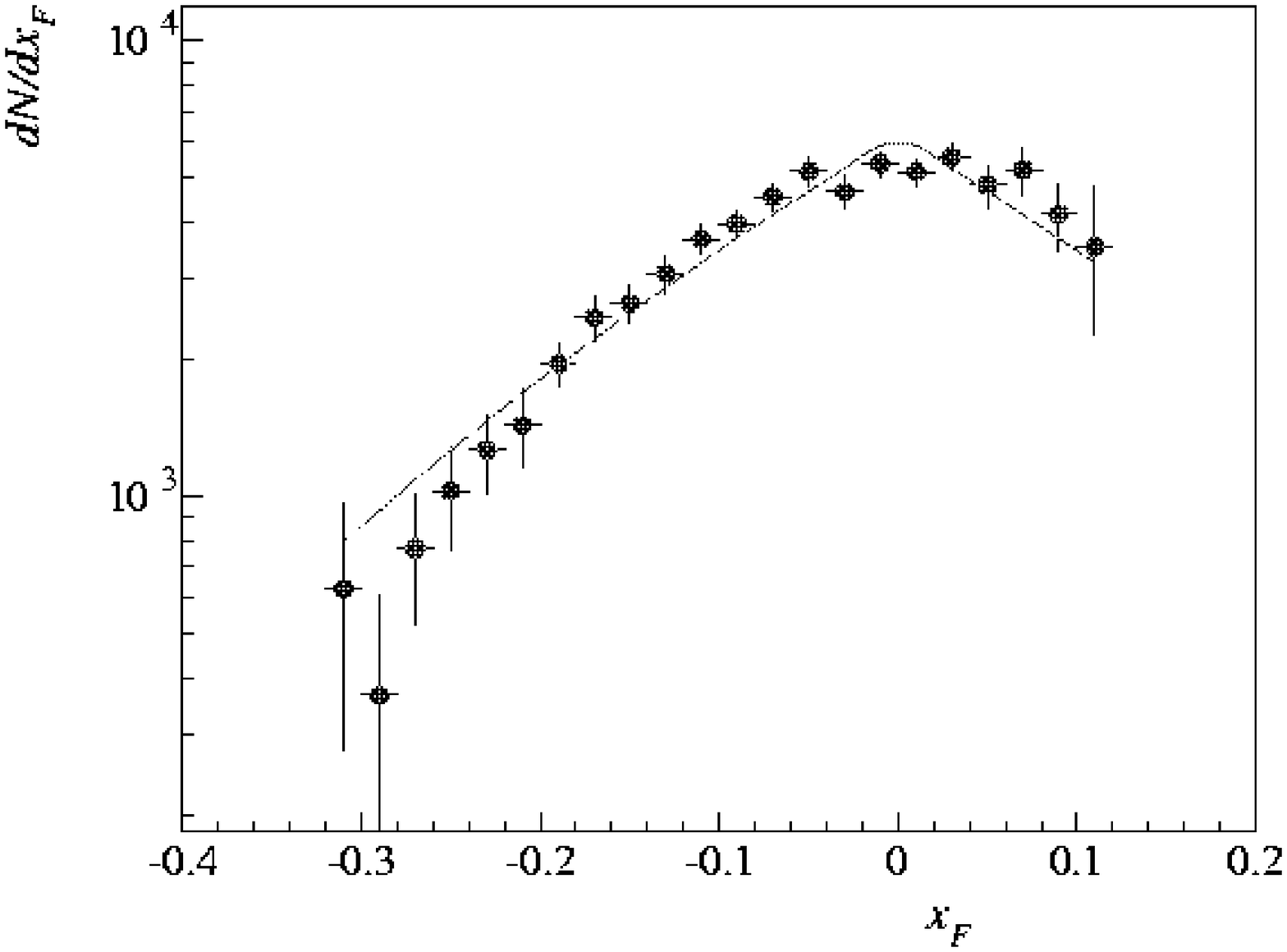,height=2.0in}
\end{center}
\caption{$J / \psi$ differential distribution versus $x_F$.
\label{fig:xf}}
\end{minipage}
\end{center}
\end{figure}

The huge sample of $J / \psi$ decays accumulated in 2002/2003 and the large acceptance covered by the HERA-B spectrometer provide a broad coverage in $p_T$, up to 4.8 GeV.
The differential distribution for $J / \psi$ as a function of transverse momentum is shown in  Fig.~\ref{fig:pt}. Superimposed on the plot is a fit to the parameterization function:

\begin{displaymath}
   \frac{d \sigma}{d p_\mathrm{T} ^2  } =  A \cdot
   \left[ 1 + \left( \frac{35 \cdot \pi \cdot p_{T} }{256 \; \cdot <p_\mathrm{T}> } \right)^2
   \right] ^{-6} 
\end{displaymath}

Our results on the average transverse momentum as well as the results~\cite{e771,e789au} from E771 and E789 experiments are shown in Table~\ref{tab:jpsipt}.
The results obtained in electron and muon channel are compatible with each other for each target material. The HERA-B measurement confirms the trend of increasing average $p_T$ with increasing atomic number and energy in the center of mass. 

The differential distribution of $J / \psi$ versus Feynman $x_F$ is plotted in Fig.~\ref{fig:xf}. 
The points and errors are shown along with a fit to the parameterization function: 

\begin{displaymath}
   \frac{d \sigma}{d x_F  } =  A \cdot ( 1 - | x_F | )^C
\end{displaymath}

The HERA-B differs from all other proton-nucleus collision experiments by its large acceptance in the negative $x_F$ region. Our measurement for exponential parameter and the results~\cite{e771,e789au,e789cu} from previous experiments are shown in Table~\ref{tab:jpsixf}. Due to the ongoing acceptance estimations for different running periods, at the moment only a range for the slope is quoted. 

\begin{table}[ht]
\caption{The HERA-B results on the average transverse momentum for $J / \psi$.
Comparison with data from E771 and E789 experiments.\label{tab:jpsipt}}
\vspace{0.4cm}
\begin{center}
\begin{tabular}{|c|c|c|c|c|c|}
\hline
Experiment &
Beam Energy &
Target &
$p_T$ Range &
$<p_{T}> (e^+e^-)$ &
$<p_{T}> (\mu^+\mu^-)$
\\ \hline
HERA-B &
920 GeV &
C &
$p_T<4.8$ GeV &
$1.22\pm0.01_{stat}$ &
$1.22\pm0.01_{stat}$ 
\\ \hline
HERA-B &
920 GeV &
W &
$p_T<4.8$ GeV &
$1.29\pm0.01_{stat}$ &
$1.30\pm0.01_{stat}$ 
\\ \hline
E771 &
800 GeV &
Si &
$p_T<3.5$ GeV &
&
$1.20\pm0.01$ 
\\ \hline
E789 &
800 GeV &
Au &
$p_T<2.6$ GeV &
&
$1.29\pm0.009$ 
\\ \hline
\end{tabular}
\end{center}
\end{table}

\begin{table}[ht]
\caption{The HERA-B results on the exponential parameter for $J / \psi$ distribution in $x_F$. Comparison with data from E771 and E789 experiments.\label{tab:jpsixf}}
\vspace{0.4cm}
\begin{center}
\begin{tabular}{|c|c|c|c|c|}
\hline
Experiment &
Beam Energy &
Target &
$x_F$ Range &
c 
\\ \hline
HERA-B &
920 GeV &
C,W &
$-0.35<x_F<0.15$ &
$(5.0-6.5)\pm0.3_{stat}$ 
\\ \hline
E771 &
800 GeV &
Si &
$-0.05<x_F<0.25$ &
$6.54\pm0.23$ 
\\ \hline
E789 &
800 GeV &
Au &
$-0.03<x_F<0.13$ &
$4.91\pm0.18$ 
\\ \hline
E789 &
800 GeV &
Cu &
$0.30<x_F<0.95$ &
$5.21\pm0.04$ 
\\ \hline
\end{tabular}
\end{center}
\end{table}

\subsection{A-Dependence for  $J / \psi$}\label{subsec:prod}
The suppression of charmonium production in heavy-ion collisions is predicted to be a signature for the formation of the quark-gluon plasma. And it is therefore of particular interest to study these mechanisms in the absence of quark-gluon plasma. 
In this context the HERA-B presents the new measurement of A-Dependence for  $J / \psi$.

Nuclear effects are usually considered in a power-law parameterization:

\begin{displaymath}
\sigma_{pA} =  \sigma_{pN} \cdot  A^{\alpha} 
\end{displaymath} 
The parameter $\alpha$ can be determined from measurements with two materials:

\begin{displaymath}
\alpha = \frac{1}{log(A_W/A_C)} \cdot log \left( \frac{N_W}{N_C}
\frac{\mathcal{L}_C}{\mathcal{L}_W}
\frac{\epsilon_C}{\epsilon_W} \right)      
\end{displaymath}
where $N$ is the number of reconstructed $J / \psi$'s, $L$ is the integrated luminosity, $\epsilon$ - is the detector and trigger efficiency and the acceptance. To reduce the systematic effects only runs with C and W wires operating simultaneously were used. 

\begin{figure}[ht] 
\begin{center}
\psfig{figure=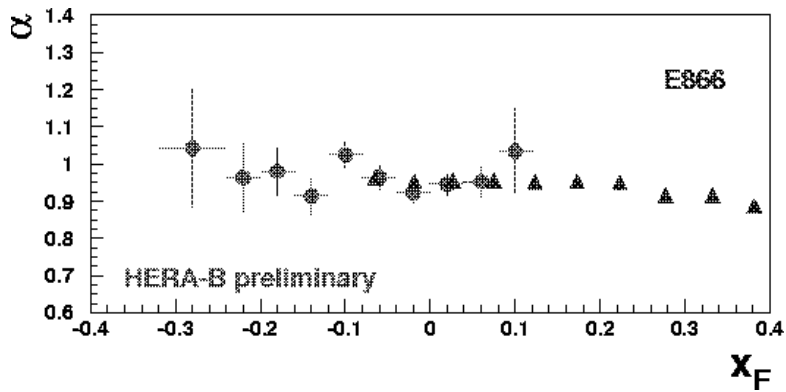,height=2.0in}
\end{center}
\caption{The A-dependence for $J / \psi$ versus $x_F$. Also shown are the results from E866 experiment.
\label{fig:alpha}}
\end{figure}

A-dependence exhibits strong kinematic dependences with Feynman $x_F$ and transverse momentum $p_T$ of the produced $J / \psi$. Therefore we plot $\alpha$ as a function of $x_F$, as shown in Fig.~\ref{fig:alpha}. Since the ratio of luminosities is under investigation at the moment, the results were normalized to the E866 data~\cite{e866}.
\subsection{Production ratio of $\psi(2S)$ to $J / \psi$}\label{subsec:prod}
\begin{figure}[ht]
\begin{center}
\hspace*{-10mm}
\begin{minipage}{2.5in}
\begin{center}
\psfig{figure=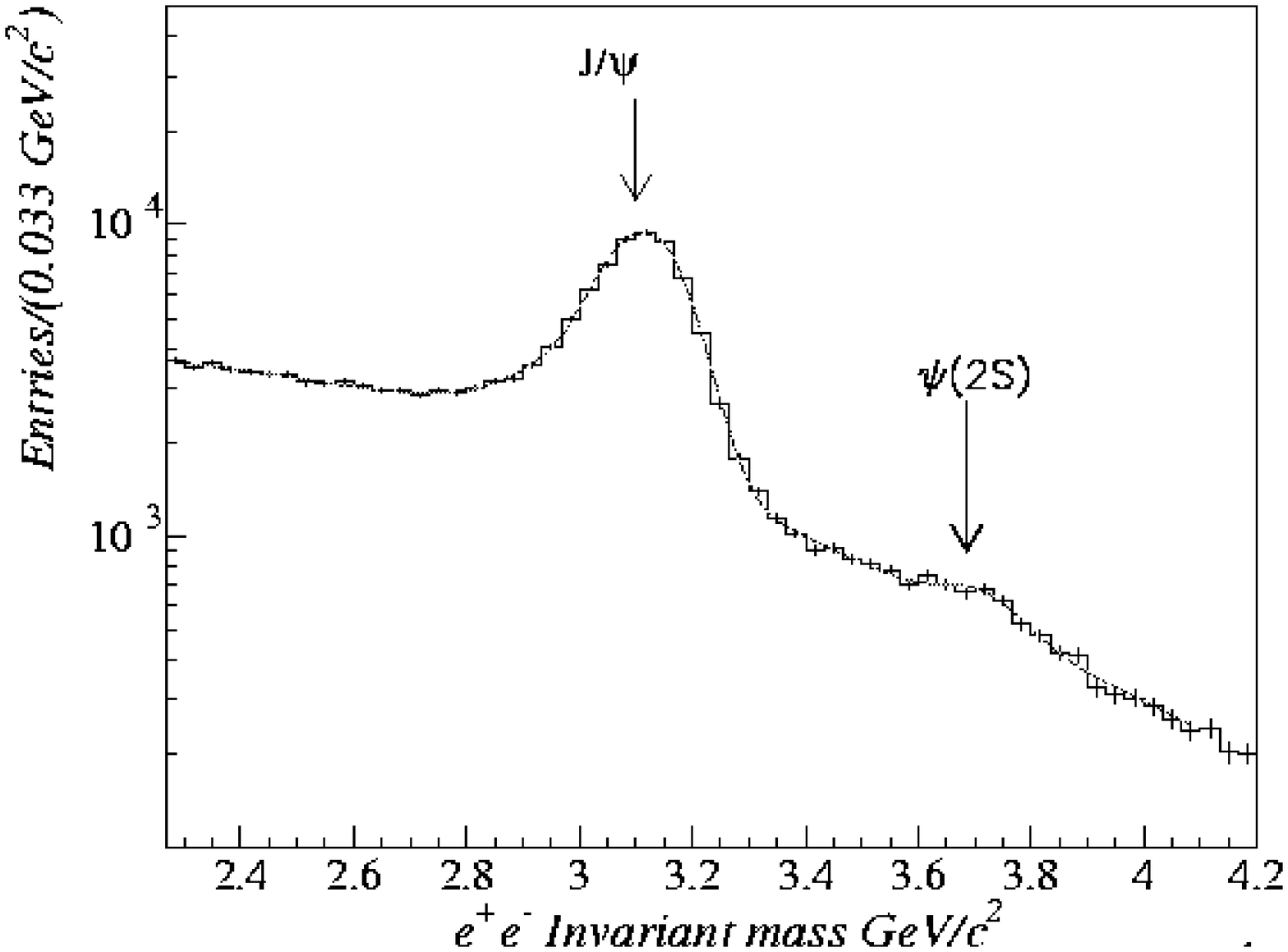,height=2.0in}
\end{center}
\caption{The $e^+e^-$ invariant mass distribution. $J / \psi$ and $\psi(2S)$ resonances are visible.
\label{fig:psiee}}
\end{minipage}
\hspace*{15mm}
\begin{minipage}{2.5in}
\begin{center}
\psfig{figure=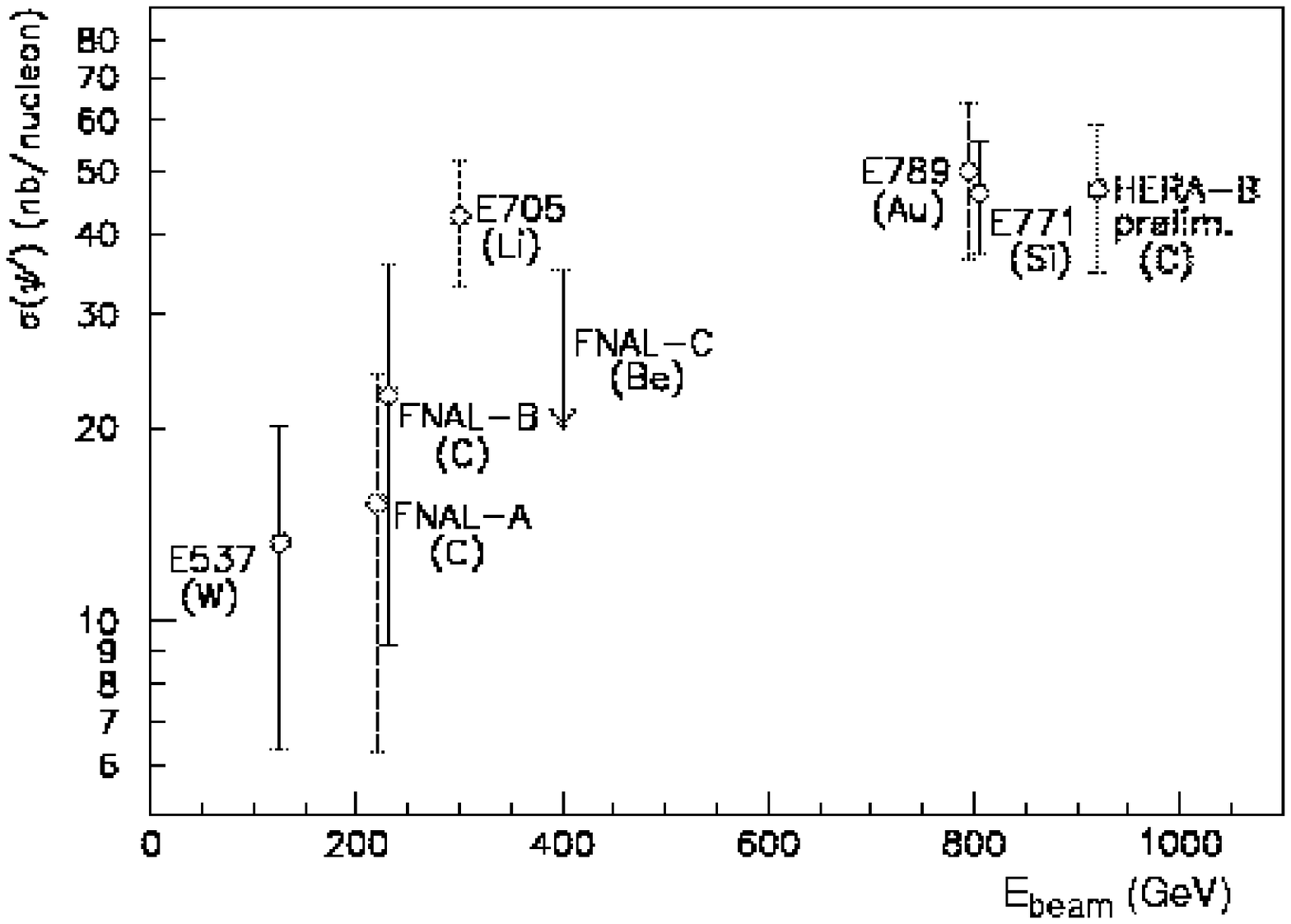,height=2.0in}
\end{center}
\caption{The proton beam energy dependence of the $\psi(2S)$ production cross section.
\label{fig:psicross}}
\end{minipage}
\end{center}
\end{figure}

The production ratio of $\psi(2S)$ to $J / \psi$ is determined according to:
\begin{displaymath}
   R_{\psi(2S)} = \frac{\sigma(\psi(2S))}{\sigma(J / \psi)} =
   \frac{N(\psi(2S))}{N(J / \psi )} \cdot
   \frac{Br(J / \psi \rightarrow l^+l^- )}{Br(\psi(2S) \rightarrow l^+l^- )} \cdot   
   \frac{\epsilon(J / \psi)}{\epsilon(\psi(2S))}
\end{displaymath}
The electron channel (see Fig.~\ref{fig:psiee}) gives $0.13\pm0.02_{stat}$ for the $R_{\psi(2S)}$ ratio, which is in good agreement with the result from the muon channel. We also measure the $\psi(2S)$ cross section relative to the $J / \psi$. The advantage of this method is that detector and trigger efficiencies largely cancel. The value for the $J / \psi$ cross section is based on E771 and E789 measurements~\cite{e771jpsicross,e789jpsicross} and corresponds to $\sigma(J / \psi) = 357\pm8_{stat}\pm27_{sys}nb/nucl$~\cite{bb}. The resulting value for the  $\psi(2S)$ cross section is $\sigma(\psi(2S)) = 46\pm12_{stat}nb/nucl$ and is shown in Fig.~\ref{fig:psicross} along with the results from previous experiments.
\subsection{Fraction of $J / \psi$ produced via $\chi_c$ decays}\label{subsec:prod}

The production ratio of $\chi_c$ to  $J / \psi$  can help to discriminate between different theoretical models. The $\chi_c$ is reconstructed in the decay $\chi_c \rightarrow J / \psi \gamma \rightarrow l^+ l^- \gamma$. 
Due to the small branching ratio of $ \chi_{c0} \rightarrow J / \psi \gamma$, only $\chi_{c1}$ and $\chi_{c2}$ contribute to the observed $\chi_c$ signal. The energy resolution in the HERA-B is insufficient to distinguish $\chi_{c1}$ and $\chi_{c2}$ states.
The fraction of $J / \psi$ produced via radiative $\chi_c$ decays is determined from:

\begin{displaymath}
  R_{ \chi_c } = \frac{ \sum_{i=1}^{2} \sigma(\chi_{ci}) \cdot Br( \chi_{ci}
  \rightarrow J / \psi \gamma) }{\sigma(J / \psi)} =
  \frac{N(\chi_c)}{N(J / \psi)} \cdot 
  \frac{\epsilon (J / \psi)}{\epsilon (\chi_c) \cdot \epsilon (\gamma ) }
\end{displaymath}

The published result based on data from an earlier HERA-B run with limited statistics for $\chi_c$ is $ R_{\chi_c} = 0.32 \pm 0.06_{stat} \pm 0.04_{sys} $~\cite{chic}, which is compatible with most of the previous data and the NRQCD prediction.

\begin{figure}[ht]
\begin{center}
\begin{minipage}{2.5in}
\begin{center}
\psfig{figure=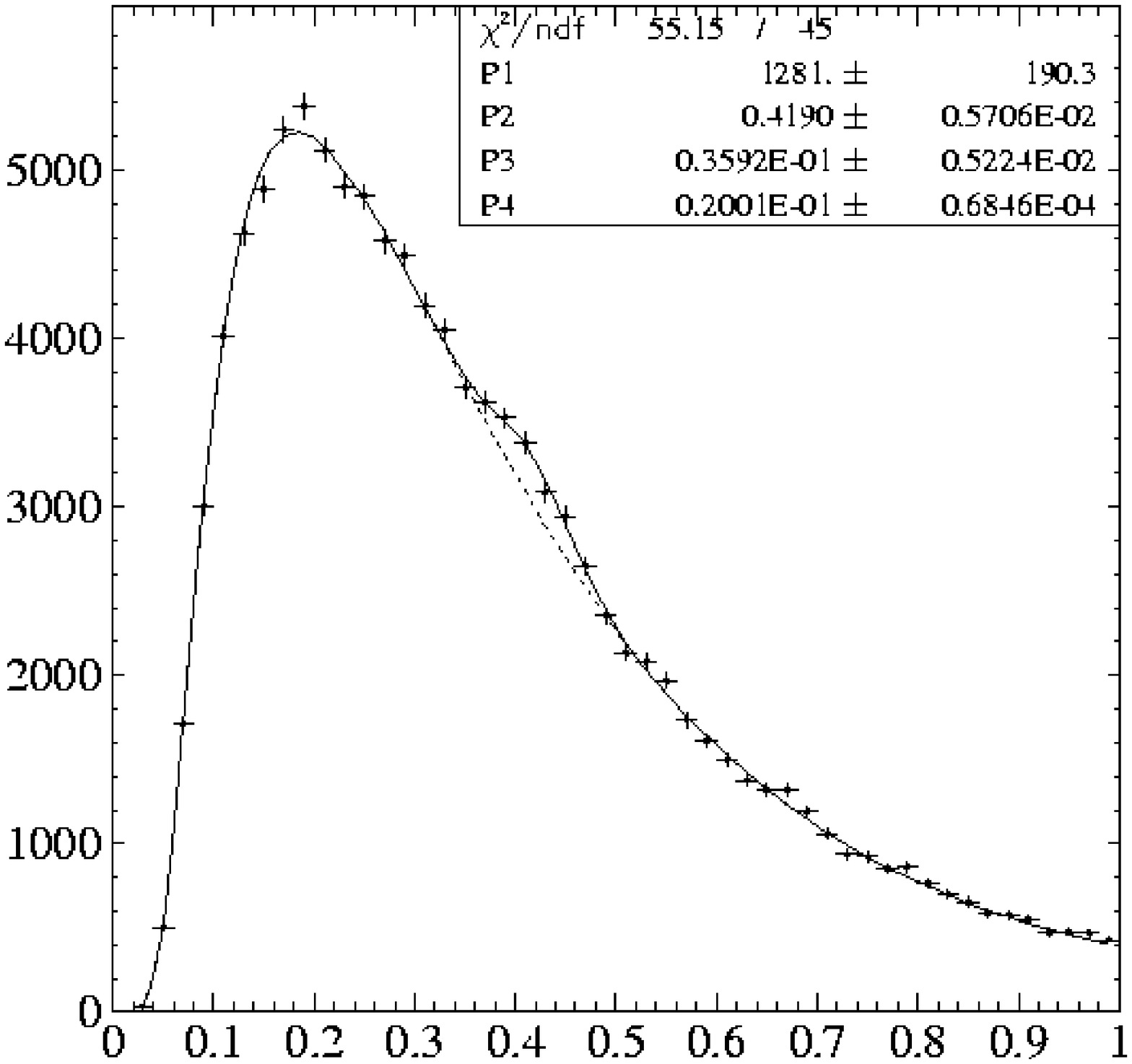,height=2.2in}
\end{center}
\caption{The $\Delta M=M(l^+l^-\gamma)-M(l^+l^-)$ distribution.
\label{fig:chic}}
\end{minipage}
\hspace*{15mm}
\begin{minipage}{2.5in}
\begin{center}
\psfig{figure=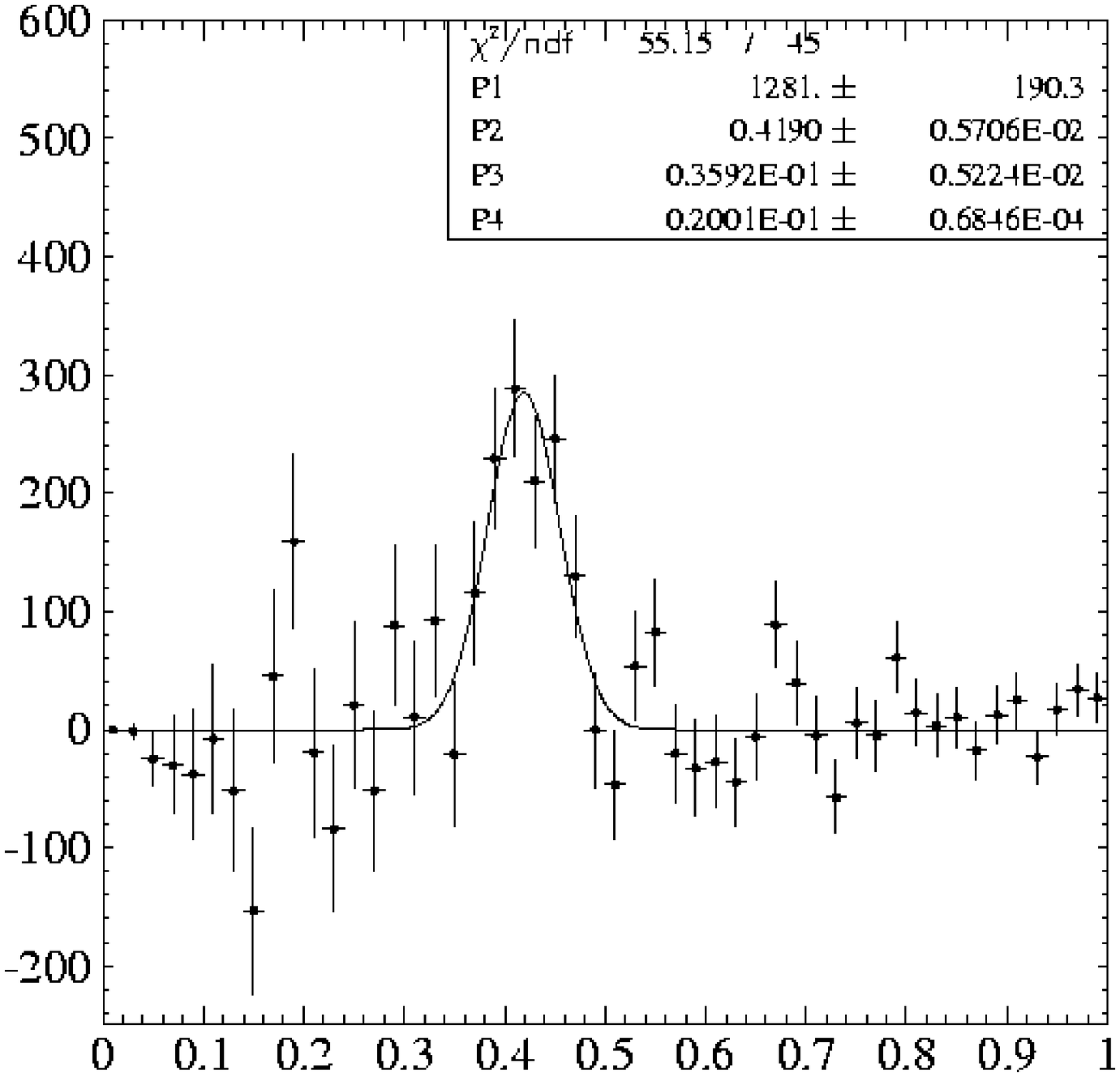,height=2.2in}
\end{center}
\caption{The $\chi_c$ signal after background subtraction.
\label{fig:chicdifference}}
\end{minipage}
\end{center}
\end{figure}

In the 2002/2003 HERA run, HERA-B collected a $\chi_c$ sample which significantly exceeds the statistics obtained by previous experiments.
The mass difference distribution $\Delta M=M(l^+l^-\gamma)-M(l^+l^-)$ is shown in Fig.~\ref{fig:chic}.  An excess of events with respect to combinatorial background determines the number of $\chi_c$ candidates. The same plot, but after background subtraction is shown in Fig.~\ref{fig:chicdifference}. 
The shape of the background is taken by combining $J / \psi$ and photon candidates from different events. The plot corresponds to $15 \%$ of statistics available in muon channel. We expect about 15,000 $\chi_c$ for the full data set combined from electron and muon samples. The preliminary result for $ R_{\chi_c}$ from the 2002/2003 data is $0.21\pm0.05_{stat}$.
\section{Summary}

HERA-B has finished data-taking and now focuses on analysis of large accumulated data samples: 150 million of events with di-lepton trigger and 210 million of events with minimum bias trigger. These data were obtained on a variety of nuclear targets. About 300,000 $J / \psi$'s decaying into electron and muon pairs were reconstructed in the di-lepton sample. HERA-B features broad coverage in $p_T$ and explores the negative $x_F$ region for the first time. Differential distributions for $J / \psi$ as functions of $p_T$ and $x_F$ are shown. New data on $J / \psi$ suppression are presented. Results on the production ratios of $\psi(2S)$ to $J / \psi$ and $\chi_c$ to $J / \psi$ are also shown. Clear signals of $D^+$, $D^0$ and $D^{*+}$ mesons were obtained in the minimum bias data. Results on $D^+$ and $D^0$ production cross sections and the $D^+$ to $D^0$ ratio are discussed.

Lots of physics analyses are ongoing and we are looking forward to new results.

\end{document}